\documentstyle[11pt]{article}

\renewcommand{\thefootnote}{\alph{footnote}}

\newcommand{\beqn}{\begin{equation}}
\newcommand{\eeqn}{\end{equation}}

\renewcommand{\thefootnote}{\alph{footnote}}

\begin{document}

\begin{titlepage}

\today          \hfill
\begin{center}
\hfill    LBNL-41006	 \\
\hfill    UCB-PTH-97/54 \\

\vskip .25in
\renewcommand{\thefootnote}{\fnsymbol{footnote}}
{\large \bf A Non-renormalization Theorem for the
Wilsonian Gauge Couplings in 
Supersymmetric Theories}\footnote{This work was supported in
part by the Director, Office of Energy Research, Office of High Energy and 
Nuclear Physics, Division of High Energy Physics of the U.S. Department of 
Energy under Contract DE-AC03-76SF00098 and in part by the National Science
Foundation under grant PHY-95-14797. MG was supported by NSERC.}
\vskip .25in

Michael Graesser\footnote{email address: graesser@thsrv.lbl.gov} and 
Bogdan Morariu\footnote{email address: bogdan@physics.berkeley.edu}
\vskip .25in

{\em 	
        Department of Physics			\\
	University of California, Berkeley,
        California 94720, USA 				\\
				and									\\
	Theoretical Physics Group			\\
	Lawrence Berkeley National Laboratory	\\
        1 Cyclotron Rd., 50A-5101
	Berkeley, California 94720, USA}
\end{center}
\vskip .25in

\begin{abstract}
We show that the holomorphic 
Wilsonian {\em beta}\/-function of 
a renormalizable 
asymptotically free supersymmetric gauge theory with an 
arbitrary semi-simple 
gauge group,
matter content, and renormalizable
superpotential
is exhausted at 1-loop with no higher loops and 
no non-perturbative contributions. This is a non-perturbative 
extension of the well known result of Shifman and Vainshtein.
\end{abstract}
\end{titlepage}
\renewcommand{\thepage}{\roman{page}}
\setcounter{page}{2}
\mbox{ }

\vskip 1in

\begin{center}
{\bf Disclaimer}
\end{center}

\vskip .2in

\begin{scriptsize}
\begin{quotation}
This document was prepared as an account of work sponsored by the United
States Government. While this document is believed to contain correct
 information, neither the United States Government nor any agency
thereof, nor The Regents of the University of California, nor any of their
employees, makes any warranty, express or implied, or assumes any legal
liability or responsibility for the accuracy, completeness, or usefulness
of any information, apparatus, product, or process disclosed, or represents
that its use would not infringe privately owned rights.  Reference herein
to any specific commercial products process, or service by its trade name,
trademark, manufacturer, or otherwise, does not necessarily constitute or
imply its endorsement, recommendation, or favoring by the United States
Government or any agency thereof, or The Regents of the University of
California.  The views and opinions of authors expressed herein do not
necessarily state or reflect those of the United States Government or any
agency thereof, or The Regents of the University of California.
\end{quotation}
\end{scriptsize}

\vskip 2in

\begin{center}
\begin{small}
{\it Lawrence Berkeley National Laboratory is an equal opportunity employer.}
\end{small}
\end{center}

\newpage
\renewcommand{\thepage}{\arabic{page}}

\setcounter{page}{1}
\setcounter{footnote}{0}

\section{Introduction}
In an important paper~\cite{shifman} Shifman and Vainshtein solved 
the anomaly puzzle in supersymmetric gauge theories by 
first introducing the Wilsonian 
gauge {\em beta}\/-function, which they proved to be 1-loop perturbatively 
exact. Then they argued that the supersymmetric extension of the anomaly 
equation should be written in operator form. Furthermore, they showed 
that the coefficient of the anomaly 
involves the Wilsonian gauge {\em beta}\/-function rather 
than the exact Gell-Mann 
and Low function~\cite{shifman2}.
In this Letter we give an extension of their work on the exactness of the 
1-loop Wilsonian {\em beta}\/-function, and show that there are
no further non-perturbative violations.
More specifically, we prove that the holomorphic 
Wilsonian {\em beta}\/-function of an arbitrary 
renormalizable asymptotically-free 
supersymmetric gauge theory with matter 
is exhausted at 1-loop with no
higher loops and no non-perturbative contributions.

The technique
we employ to prove the theorem
was introduced by Seiberg~\cite{seiberg} and it is briefly 
reviewed here. 
To obtain the {\em beta}\/-function we compare two versions of the theory 
with different cutoffs and coupling constants and the same low energy
physics. The couplings of the theory
with the lower cutoff can be expressed in terms of the couplings of the
theory with the higher cutoff and the ratio of the two cutoffs. We can
restrict their functional dependence  on the high cutoff couplings 
using {\em holomorphy} of the
superpotential and gauge kinetic terms and {\em selection rules}. 
Holomorphy is a consequence of supersymmetry. To see this, 
elevate the couplings to
background chiral superfields. They must appear holomorphically in the 
superpotential in order to preserve supersymmetry. Selection rules generalize
global symmetries in the sense that we allow the couplings in the
superpotential to transform under these symmetries.
Non-zero vacuum values of these couplings then spontaneously break these
symmetries. Here we only consider $U(1)$\/ and $U(1)_R$\/ symmetries. 
In the quantum theory they are generally anomalous, but we can use the same
technique we used for the coupling in the superpotential. We assume that the
$\theta$\/-angle is a background field and transform it non-linearly
to make the full quantum effective action invariant. 

Then, following
a method used in~\cite{nima} we translate
these conditions on the functional relations between the couplings 
of the theories at different cutoffs into
restrictions of the functional form of the 
gauge {\em beta}\/-function. We can show that the gauge 
{\em beta}\/-function
is a function of the holomorphic invariants allowed by selection rules. 
Then we can restrict further the functional dependence of 
the {\em beta}\/-function by varying the couplings while keeping the 
invariants fixed.  This allows us to 
relate the {\em beta}\/-function of the original
theory to the {\em beta}\/-function of a 
theory with vanishing superpotential.
In addition, we also obtain a strong restriction of the functional dependence 
of the {\em beta}\/-function on the
gauge coupling. It has exactly the form of a one-loop 
{\em beta}\/-function. 
The only ambiguity left is a numerical coefficient which can be calculated
in  perturbation theory.

Next we make a short detour to explain what we mean by the Wilsonian
{\em beta}\/-function~\cite{polchinski}. 
The Wilsonian
{\em beta}\/-function describes the renormalization group flow of the 
{\em bare}
couplings of the theory so that the low energy theory is cutoff invariant.
Additionally, we do not renormalize the vector and chiral superfields,
i.e. we do not require canonical normalization
of the kinetic terms~\cite{shifman}.
The usual convention in particle physics is to canonically
normalize the 
kinetic term. It is obtained by using the covariant derivative
$\partial + g A$\/. Instead, here we allow non-canonical normalization
of the kinetic term. The normalization of the gauge fields is such that
the covariant derivative has the form
$\partial + A$\/. The gauge coupling only appears in front of the gauge
kinetic term. In this case it is convenient to
combine the $\theta$\/-angle and gauge coupling constant  $g$\/
into the complex variable $\tau = \theta/2 \pi + 
4 \pi i/g^2$. In supersymmetric gauge theories the 
{\em beta}\/-function is holomorphic in the bare couplings only if we do 
not renormalize the fields.
Even if we start with canonical normalization at a higher cutoff, the Kahler
potential will not be canonical at the lower cutoff. 
The rescaling of the chiral
or gauge superfields 
is an anomalous
transformation~\cite{shifman} that destroys the 
holomorphy of the superpotential and
the {\em beta}\/-function
\footnote{For some special theories like $N=2$\/ SUSY-YM
the rescaling anomaly of the chiral superfields cancels 
the rescaling anomaly
of the vector superfield~\cite{nima}. For these 
theories we can make stronger
statements since the canonical and holomorphic Wilsonian couplings coincide.}. 
The relation between the 
{\em beta}\/-functions in the two
normalizations was first discussed in~\cite{shifman}. 
The {\em beta}\/-function for
canonically normalized fields is known exactly~\cite{shifman2} and 
receives contributions to all orders in 
perturbation theory. For a recent discussion of these issues  see 
also~\cite{nima}.
Again we emphasize that 
here we are only concerned with the holomorphic Wilsonian 
{\em beta}\/-function.

 We should also clearly state that the theorem 
is not valid if any one of the one-loop gauge 
{\em beta}\/-functions is not asymptotically free. This includes the case
when the one-loop {\em beta}\/-function vanishes. As we will see, exactly
in this case the $U(1)_R$\/ symmetry is non-anomalous. This makes it 
difficult 
to control the dependence of the {\em beta}\/-function on the gauge coupling.

Various partial versions 
of this result already existed. As we already mentioned, the perturbative
non-renormalization theorem was proven in~\cite{shifman}.
An analysis of possible non-perturbative 
violations to this theorem in the case of a simple 
gauge group with a vanishing  
superpotential could be found in~\cite{nima}.
It was also known that in the case
of a simple gauge group with only Yukawa interactions 
present in 
the superpotential, possible non-perturbative corrections to the 
Wilsonian {\em beta}\/-function are independent of 
the gauge coupling~\cite{nima2}. 

Finally, we note that the theorem is valid in theories where no mass terms 
are allowed by the symmetries of the theory. This is of 
phenomenological interest as many supersymmetric 
extensions of the Standard Model share this characteristic.

\section{Simple Gauge Group}
\label{simpleGG}
We will consider first the case of a simple gauge group $G$\/.
Let the generalized superpotential 
$\widetilde{W}$\/ be defined to include the 
kinetic term for the gauge fields
\begin{equation}
 \widetilde{W} = \frac{\tau}{64 \pi i t_R } 
\hbox{tr}_R(W_{\alpha}W^{\alpha})   + W,
\end{equation}
where
\begin{equation}
 W = 
\sum \lambda_{ijk} \Phi_{i}\Phi_{j}\Phi_{k} +
 M \sum m_{ij} \Phi_{i} \Phi_{j} + M^{2} \sum c_{i} \Phi_{i} 
\end{equation}
is the usual superpotential and 
$\hbox{tr}_R T^a T^b=t_R \delta^{ab}$.
Here $M$\/ is the cutoff mass and was factored 
out so that all the couplings are
dimensionless. The gauge 
coupling $g$\/ and $\theta$-angle are combined 
in the complex variable
\begin{equation}
	\tau \equiv \frac{\theta}{2 \pi} + 
\frac{4 \pi i}{g^2}.
\end{equation}	
Note that unitarity requires 
$\tau$\/ to be valued in the upper half plane.
Since $\theta$\/ is a periodic variable it is 
convenient to introduce a new
variable $q \equiv e^{2\pi i \tau}$\/. It is valued in the 
complex plane and 
transforms linearly under the anomalous 
transformations to be discussed 
below. Weak coupling is at $q=0$\/.

Consider now a theory with a different cutoff 
$M'$\/ and with the same low
energy physics. The Lagrangian at the new cutoff is 
\begin{equation}
 {\cal L}=\sum_{i} \int d^2 \theta d^2 \bar{\theta} 
Z_i \Phi_{i}^{\dagger} e^{2V_h}
\Phi_i +\left(\int d^2 \theta \widetilde{W}(\tau',\lambda'_{ijk},
m'_{ij},c'_{i}, M')+\hbox{h.c.}
\right) 
\end{equation} 
where
in particular, the fields $\Phi_i$ are not renormalized to 
canonical normalization. The $Z_i$\/ depends 
non-holomorphically on the couplings, so renormalizing the chiral
superfields would destroy the holomorphic form of $\widetilde{W}$\/.
The new coupling $\tau'$\/ is a function of the old
dimensionless couplings and the 
ratio $M/M'$\/. For later convenience we 
write this as
\begin{equation}
	\tau' = 
\tau' ( \tau, \lambda_{ijk},m_{ij},c_{i};\ln (M/M') ).
\label{tau}
\end{equation}
 Supersymmetry requires a holomorphic 
dependence of $\tau$\/ on the first 
four arguments. To see this, note that the couplings
in the generalized superpotential 
can be considered as vacuum values of
background chiral superfields.  Invariance of the action under
supersymmetry transformations 
requires holomorphy of the superpotential.

To prove the non-renormalization 
theorem we will use selection rules. These
are global symmetries of the superpotential 
with all couplings considered as 
chiral superfields. We assign them 
non-trivial transformation properties 
under the symmetry group. These symmetries will be 
spontaneously broken by non-zero 
vacuum values of the couplings.
In general they are also anomalous. We 
will make them non-anomalous by 
assigning a charge to $q$\/, i.e. 
transforming $\theta$\/ to compensate for 
the anomaly. Consider the 
$U(1)_{R}\times U(1)$\/ global symmetry with the 
following charge assignment:

\[	\begin{array}{|c|c|c|c|c|c|c|}
\hline
		  
&W_{\alpha}&\Phi_{i}&\lambda_{ijk}&m_{ij}&c_{i}&  q       \\
\hline
U(1)_{R}&	1		 &	2/3	 &	       
0		& 2/3	 & 4/3 & 2 b_{0}/3    \\
\hline
U(1)	  &	0		 &	
1      &       -3  	&	-2  &	-1  &2\sum_{i} t(R_i)\\
\hline
	\end{array}		\]

The quantity $b_{0}$ is given 
by $b_{0}=3t_{adj}-\sum_{i} t(R_i)$,
where $t(R_i)$ is the normalization 
of the generators for the representation of the
chiral superfield $\Phi_i$\/. For example, 
$t=1/2$ for a fundamental 
of $SU(N)$. 
Define the gauge $\beta$\/-function by
\beqn
\beta_{2\pi i \tau} =  \frac{d}{d \ln(M/M')} 2\pi i \tau' |_{M'=M} =
\beta ( \tau , \lambda_{ijk} , m_{ij}, c_{i} ).
\label{beta}
\eeqn
The holomorphy of $\tau$\/ in~(\ref{tau}) translates into 
holomorphy
of the {\em beta}\/-function.

Since $\tau \rightarrow \tau + 1$\/ 
is a symmetry of the theory, $\beta$\/ is
a single valued function of $q$
\beqn
\beta_{2 \pi i \tau} = f(q,\lambda_{ijk} , m_{ij}, c_{i} ).
\eeqn
First, consider the case when at 
least one mass term, let us call it
 $m_{*}$\/, can 
be non-zero. If any $c_i$\/ could be non-zero, then there is a gauge
singlet field which could be given a Majorana mass, so this is the same 
case as above.

The gauge {\em beta}\/-function is $U(1)_{R} \times U(1)$\/ invariant.
This statement is non-trivial and requires some explanation. 
Consider some arbitrary coupling $\lambda$\/ that transforms linearly 
under some $U(1)$\/ or $U(1)_R$\/ symmetry. Its 
{\em beta}\/-function $\beta_{\lambda}$\/ must also transform linearly 
with the same charge as $\lambda$\/
\beqn
e^{i Q_{\lambda} \alpha} \beta_{\lambda}(\lambda,\ldots) = 
\beta_{\lambda}(e^{i Q_{\lambda} \alpha}\lambda,\ldots)
\eeqn
where $Q_{\lambda}$\/ is the charge of $\lambda$\/. This is true
in particular for the {\em beta}\/-function of $q$\/. However
when we go to the $\tau$\/ variable we have
\beqn
\beta_{2 \pi i \tau} = \frac{d}{d \ln (M/M')} 2 \pi i  \tau' = 
\frac{d}{d q}2 \pi i \tau' ~\beta_{q} = q^{-1} ~\beta_{q}.
\eeqn
The additional $q$\/ factor makes the $\tau$\/ {\em beta}\/-function
invariant. In what follows we only consider the gauge {\em beta}\/-function
since all the others are trivial, i.e. there are no 
perturbative \cite{nonrenorm} or non-perturbative
\cite{seiberg} corrections to the usual superpotential.
We will drop the subscript and denote it 
$\beta$\/.

First consider $U(1)_R$\/ invariance. It requires that
\beqn
\beta = \widetilde{f} \left(\frac{q}{m_{*}^{b_0}},
\frac{m_{ij}}{m_{*}},
\frac{c_i}{m_{*}^{2}} , \lambda_{ijk} \right).
\eeqn
However, the variables of 
$\widetilde{f}$\/ are not $U(1)$\/ invariant. They
have charges $6 T_{adj}, 0,3,-3$\/, respectively. Invariance under 
$U(1)_{R} \times U(1)$\/ requires 
that $\beta$\/ is a yet another function 
\beqn
\beta = F \left(
q~\frac{\lambda_{ijk}^{2 t_{adj}}}{m_{*}^{b_0}}, 
q^{-1} ~\frac{{c_i}^{2t_{adj}}}{  m_{*}^{t_{adj}+\sum t_i}}, 
\frac{m_{ij}}{m_{*}} 
\right).
\eeqn
We next take the limit $m_* \rightarrow 0$ keeping $q$\/ 
and all the arguments
of $F$\/ constant. 
If $b_0 > 0$\/, this corresponds to taking
all couplings except $\tau$\/ to zero. Assuming that
$\beta$\/ is continuous we see that  
$\beta(q,\lambda_{ijk} , m_{ij}, c_{i})=\beta(q,
\lambda_{ijk} = m_{ij}=c_{i}=0)$\/ and thus it is
independent of all the couplings in the 
superpotential. In fact when the 
superpotential
vanishes it is known~\cite{nima} that there are no non-perturbative 
corrections to the 
{\em beta}\/-function and the 
gauge coupling only runs at 
1-loop\footnote{Note that this result can also be 
written  as $\frac{d}{dt} g 
= - \frac{b_0}{16 \pi^2} g^{3}$\/ which is just 
the standard 1-loop {\em beta}\/-function.}. 
This just reflects the fact that no $U(1)_R \times U(1)$\/ 
holomorphic invariant can be constructed solely in terms of $q$\/.
Note the importance of holomorphy in these arguments. For 
example, if we do not require holomorphy $q \bar{q}$\/ is
invariant under an arbitrary $U(1)$\/ and $U(1)_R$ symmetry.
No higher loops or 
non-perturbative corrections are present and we conclude
that
\beqn
\beta = b_{0},
\eeqn
thus extending the perturbative result of Shifman and 
Vainshtein~\cite{shifman}.

An exception to the previous argument occurs when the gauge and
global symmetries of the theory allow only 
Yukawa couplings to be present in the superpotential. 
For these theories  
\beqn
\beta=f(q,\lambda_{ijk}). 
\eeqn
The {\em beta}\/-function must be $U(1)_R$\/ invariant. This requires
\begin{equation}
f(e^{2b_0 \alpha i/3}q,\lambda_{ijk})=f(q,\lambda_{ijk}).
\nonumber
\end{equation}
Then by holomorphy $\beta$\/ is independent of $q$.
Further, invariance of
$\beta$\/ under the $U(1)$ symmetry requires
that $f$ is a function of ratios of $\lambda_{ijk}$ only.
We may choose one of the non-zero $\lambda_{ijk}$,
$\lambda_{*}$ say, and divide through by $\lambda_{*}$. 
Then
\beqn
 \beta=f(\lambda_{ijk}) \
=F\left(\frac{\lambda_{ijk}}{\lambda_{*}}\right).
\eeqn 
Consider the limit $\lambda_{ijk}\rightarrow0$ while
keeping the ratios $\lambda_{ijk}/\lambda_{*}$ constant. We know
that in this limit $\beta$\/ reduces to the one-loop result.
So assuming that $\beta$\/ is continuous, we find
$\beta(\lambda_{ijk})=\beta(\lambda_{ijk}=0)=b_0$\/, i.e. it
is independent of the Yukawa couplings.

To conclude this Section, we note that 
our discussion of the proof of the theorem was divided
into two cases requiring separate proofs. Here we present 
a short argument that extends the proof of the theorem, valid
when at least one mass term is allowed, to theories which do not
admit any bare mass terms.
Consider a theory with Lagrangian ${\cal L}$\/
for which the symmetries of the theory 
forbid the presence of any mass terms.
To this theory, ${\it add}$ a 
non-interacting gauge-singlet field with 
mass $m_{*}$. 
More concretely, the new 
theory defined at $M$ is described by the Lagrangian
\begin{equation}
{\cal L}_{new}={\cal L} +\int d^2 
\theta d^2 \bar{\theta} \Phi^{\dagger}_0 \Phi_0+ 
\left(\int d^2 \theta 
M m_{*} \Phi^2_0 +\hbox{h.c.} \right). \nonumber
\end{equation} 
This new theory satisfies the conditions of the theorem 
proven when at least one mass term is allowed, 
so the {\em beta}\/-function of the new theory, $\beta_{new}$\/, 
is exhausted at one-loop. But we can conclude on 
physical grounds that $\beta_{new}$\/ is identical to $\beta$\/, 
the {\em beta}\/-function of the original theory, since in integrating 
over momentum modes $M$ to $M'$ the contribution from the 
gauge singlet completely factors out since it is non-interacting.
So by this argument the proof of the theorem for theories 
with mass terms can be extended to theories for which mass 
terms are forbidden by the symmetries of the model.  

The results of this section are also valid 
for a semisimple gauge group. We shall sketch
the proof in the next section.

\section{Extension to a semi-simple gauge group}

Assume that the gauge group is $G=\Pi_{A} G_A$ with
each
$G_A$ a simple group. Also assume that the
superpotential has the form given in Section \ref{simpleGG}.
Then if all the simple
gauge groups are asymptotically-free
the Wilsonian {\em beta}\/-functions of all the gauge couplings
are one-loop exact.

For each simple gauge group $G_A$ define
\beqn
 \tau_A=\frac{\theta_A}{2\pi}+\frac{4\pi i}{g^2_A} 
\eeqn
and introduce $q_A\equiv e^{2\pi i \tau_A}$ as in
Section \ref{simpleGG}. We extend 
the $U(1)_R \times U(1)$ selection rules of
Section \ref{simpleGG} by assigning all gauge chiral multiplets
$W_{{\alpha},A}$\/ charge $(1,0)$\/.
Then $q_A$ has charge $(2b^A_0/3,2 \sum_{i} t_A(R_i))$. It will be
conveinent to define $\kappa_A\equiv (q_A)^{\frac{1}{b^A_0}}$.
Then $\kappa_A$
has charge $(2/3, 2 \sum_{i} t_A(R_i)/ b^A_0)$. Weak coupling is
at $\kappa_A=0$ since $b^A_0$ is positive.

The {\em beta}\/-functions for each simple gauge group are defined
as in Section \ref{simpleGG}, so that
\beqn
 \beta_A=f_A(q_B,\lambda_{ijk},m_{ij},c_i) 
\eeqn
is a function of holomorphic invariants and 
invariant under the $U(1)_R \times U(1)$ symmetry.

We do the proof for two cases: 
\begin{enumerate}
   \item Only Yukawa couplings are allowed.
   \item At least one $m_{ij}\neq 0$\/ is allowed.
\end{enumerate}

In the first case invariance of $\beta_A$\/ under
$U(1)_R$ requires that $\beta_A$\/ is a function of
ratios of $\kappa_B$ only.
That is,
\beqn
 \beta_A=F_A(\kappa_B/\kappa_{B_{*}},\lambda_{ijk}). 
\eeqn
We have divided through by an arbitrarily chosen
$\kappa_{B_{*}}$, so that each $\kappa_B$ other than
$\kappa_{B_{*}}$ appears in the argument of $F$ only
once. Now consider the weak coupling limit
$\kappa_B \rightarrow 0$ for all the gauge couplings.
The argument of the {\em beta}\/-functions is
\beqn
\kappa_B/\kappa_{B_{*}}=\exp 2\pi i (\tau_B/b_0^B-
\tau_{B_{*}}/b_0^{B_{*}}).
\eeqn
Since by assumption
the one-loop
{\em beta}\/-functions all have the same sign
it is possible to take this limit while
keeping the ratios $\kappa_B/\kappa_{B_{*}}$ fixed.
In this limit
the {\em beta}\/-function
is a function of the Yukawa couplings only. So assuming 
that the
{\em beta}\/-functions are continuous in this limit, we find 
that $\beta_A(\kappa_B,\lambda_{ijk})=
\beta_A(\kappa_B=0,\lambda_{ijk})=F_A(\lambda_{ijk})$.
But now we may use the $U(1)$ symmetry to conclude that 
$\beta_A$ is a function of $\lambda_{ijk}/\lambda_{*}$. The
argument of 
Section \ref{simpleGG} may now be repeated and we conclude 
that $\beta_A(q_B,\lambda_{ijk})=$ constant.

For the second case a straightforward generalization of the
argument of Section \ref{simpleGG} may be repeated and
we conclude that
\beqn \beta_A=F_A \left(\frac{\kappa_B}{\kappa_{B_{*}}}\right). 
\eeqn
Then the  argument used in the first case of this Section 
is used to conclude 
that $F_A$ is independent of all of the $q_B$ 
and superpotential couplings.
\section*{Acknowledgements}

The authors would like to thank  
Nima Arkani-Hamed, Hitoshi Murayama, Mahiko Suzuki 
and Bruno Zumino 
for many useful discussions and valuable 
comments. 
This work was supported in 
part by the Director, 
Office of Energy Research, Office of High Energy and 
Nuclear Physics, Division of High Energy Physics of the U.S. Department of 
Energy under Contract 
DE-AC03-76SF00098 and in part by the National Science 
Foundation under grant PHY-90-14797. MG was also supported 
by NSERC.

\section*{Note}
The statement of this 
theorem for the case of a simple gauge group was also 
made in the lecture 
notes~\cite{argyres}. In that 
proof the author considers 
a superpotential containing no composite operators,
i.e. only operators linear in the fundamental 
fields. Of course such superpotential is not gauge invariant.
However it is is only used in an intermediate step
to simplify the study the charge assignment for the couplings in the 
physical gauge invariant superpotential.
The $U(1)$\/
charge of the coupling of a composite operator equals the sum
of the charges of the couplings of the fundamental fields entering the 
composite. However in~\cite{argyres} it is also assumed
that the $U(1)_R$\/ charge 
of the couplings of composite gauge 
invariant operators in the 
superpotential equals the sum of the charges of 
the couplings of fundamental 
fields forming the composite. While this is true 
for usual $U(1)$\/ symmetries 
since the superpotential has charge zero and the sum of charges
of the couplings must equal minus the sum of charges of the fields
entering the composite, for $U(1)_R$\/ symmetries the superpotential has 
charge two and the arithmetic is more complicated.
Because of this, the 
proof in~\cite{argyres} only works for a superpotential
linear in matter fields, i.e. when only gauge singlet chiral superfields
are present.
We also generalized the theorem to a semi-simple
gauge group.

\end{document}